\documentclass[aps,pre,twocolumn,superscriptaddress,showpacs,showkeys]{revtex4}

\usepackage{natbib}
\usepackage{dcolumn}
\usepackage{amsmath}
\usepackage{amssymb}
\usepackage{graphicx}
\usepackage{rotating}
\usepackage{multirow}

\begin{document}

\title{Community detection algorithms: a comparative analysis}

\author{Andrea Lancichinetti}
\affiliation{Complex Networks and Systems,
Institute for Scientific Interchange (ISI), Viale S. Severo 65, 10133, Torino, Italy}
\affiliation{Physics Department, Politecnico di Torino,
Corso Duca degli Abruzzi 24, 10129 Torino, Italy}
\author{Santo Fortunato}
\affiliation{Complex Networks and Systems,
Institute for Scientific Interchange (ISI), Viale S. Severo 65, 10133, Torino, Italy}

\begin{abstract}

Uncovering the community structure exhibited by real networks is a crucial step towards an understanding
of complex systems that goes beyond the local organization of their constituents. Many algorithms have been proposed
so far, but none of them has been subjected to strict tests to evaluate their performance. Most of the sporadic tests 
performed so far involved small networks with known community structure and/or artificial graphs with
a simplified structure, which is very uncommon in real systems. Here we test several methods against a 
recently introduced class of benchmark graphs, with heterogeneous distributions of degree and community size.
The methods are also tested against the benchmark by Girvan and Newman and on random graphs.
As a result of our analysis, three recent algorithms introduced by Rosvall and Bergstrom, Blondel et al. and 
Ronhovde and Nussinov, respectively, have an excellent performance, with the additional advantage of low computational 
complexity, which enables one to analyze large systems.
 
\end{abstract}

\maketitle

\section{Introduction}

The modern science of networks is probably the most active field within the new interdisciplinary 
science of complex systems. Many complex systems can be represented as networks,  
where the elementary parts of a system and their mutual interactions are nodes and links, respectively~\cite{newman03,boccaletti06}.
Complex systems are usually organized in compartments, which have their own role and/or function. 
In the network representation, such compartments appear as sets of nodes with a high density of internal links, 
whereas links between compartments have a comparatively lower density.  
These subgraphs are called communities, or modules, and occur in a wide variety of networked systems~\cite{girvan02,fortunato09b}.

Finding compartments may shed light on the organization of complex systems and on their function. Therefore
detecting communities in networks has become a fundamental problem in network science. 
Many methods have been developed, using tools and techniques from disciplines like physics, biology, applied mathematics, 
computer and social sciences. However, it is still not clear 
which algorithms are reliable and shall be used in applications. 
The question of the reliability itself is tricky, as it requires shared definitions of community and partition
which are, at present, still missing. This essentially means that, despite the huge literature on the topic,
there is still no agreement among scholars on what a network with communities looks like. 
Nevertheless, there has been a silent acceptance of a simple network model, the {\it planted $\ell$-partition model}~\cite{condon01},
which is often used in the literature in various versions. In this model one ``plants'' a partition, consisting of a certain
number of groups of nodes. Each node has a probability $p_{in}$ of being connected to nodes of its group
and a probability $p_{out}$ of being connected to nodes of different groups. As long as $p_{in}>p_{out}$ the groups are communities,
whereas when $p_{in}\leq p_{out}$ the network is essentially a random graph, without community structure.
The most popular version of the planted $\ell$-partition model was proposed by Girvan and Newman (GN benchmark)~\cite{girvan02}. Here the graph
consists of $128$ nodes, each with expected degree $16$, which are divided into four groups of $32$. The GN benchmark 
is regularly used to test algorithms for community detection. Indeed, algorithms can be compared based on their performance
on this benchmark. This has been done by Danon et al.~\cite{danon05}. However, the GN benchmark has two drawbacks:
1) all nodes have the same
expected degree; 2) all communities have equal size. These features are unrealistic, as complex networks are known to be characterized by
heterogeneous distributions of degree~\cite{albert99,newman03,boccaletti06} and community sizes~\cite{palla05,guimera03,danon07,clauset04,lancichinetti09}.
In recent papers~\cite{lancichinetti08,lancichinetti09b}, we have introduced a new class of benchmark graphs (LFR benchmark), 
that generalize the GN benchmark by introducing power law distributions of degree and community size. 
The new graphs are a real generalization, in that the GN benchmark is recovered in the limit case in which
the exponents of the distributions of degree and community sizes go to infinity.
Most community detection algorithms perform very well on the GN benchmark, due to the simplicity of its structure.
The LFR benchmark, instead, poses a much harder test to algorithms, and makes it easier to disclose their limits.
Moreover, the LFR benchmark graphs can be built very quickly: the complexity of the construction algorithms is linear in the number of 
links of the graph, so one can perform tests on very large systems, provided the method at study is fast enough to analyze them.

For these reasons, we believe that a serious assessment of the goodness of community detection algorithms can be 
made by evaluating their performance on the LFR benchmark. In this paper we propose a comparative analysis of this kind.
After explaining briefly the LFR benchmark and how to compare partitions quantitatively
we will pass to the description the algorithms that we examined. We will present the analysis of the algorithms'
performance first on the GN benchmark and then on the LFR benchmark, in its various versions
including weighted and directed graphs, along with graphs with overlapping communities. Finally we will consider the issue of whether the
algorithms are able to give a null result, i. e. how they handle networks without expected community structure, like random graphs. 
Our analysis will reveal that there are, at present, algorithms which are fast and reliable in many situations. We will conclude 
with a summary of our results and their consequences.

\section{The LFR benchmark}
\label{sec1}

The LFR benchmark~\cite{lancichinetti08,lancichinetti09b} is a special case of the planted $\ell$-partition model, in which groups are of different sizes and 
nodes have different degrees. The node degrees are distributed according to a power law with exponent $\tau_1$; the community
sizes also obey a power law distribution, with exponent $\tau_2$. In the following, $N$ indicates the number of nodes of the network.
In the construction of the benchmark graphs, each node receives its degree once and for all, and keeps it fixed 
until the end. In this way, the two parameters $p_{in}$ and $p_{out}$ of the planted $\ell$-partition model in this case are not independent.
Once the value of $p_{in}$ is set one obtains the value of $p_{out}$ and viceversa. It is more practical to choose as independent
parameter the {\it mixing parameter} $\mu$, which expresses the ratio between the external degree of a node with respect to its community
and the total degree of the node. Of course, in general one may take different values for the mixing parameter for different nodes, but we will
assume, for simplicity, that $\mu$ is the same for all nodes, consistently with the standard hypotheses of the planted $\ell$-partition model.
By construction, the groups are communities when $p_{in}>p_{out}$. This condition can be translated into a condition on
the mixing parameter $\mu$. Let us label $k_{i}^{in}$ and $k_{i}^{out}$ the internal and external degree of node $i$ with respect to its community (which we denote with $c$).
By definition, $k_{i}^{in}$ is the number of neighbors of $i$ that belong to its community $c$ and $k_{i}^{out}$ the number of neighbors 
of $i$ that belong to the other communities. The number of available connections $k_c^{out}$ ($k_c^{in}$) outside (inside) $c$ is given by the sum of the
degrees of the nodes outside (inside) the community. If the numbers of nodes inside and outside $c$ are not too small, the sum of their degrees can be approximated by
the product of the average degree $\langle k\rangle$ by the number of nodes. We indicate with $n_c$ the number of nodes of the community $c$ of node $i$, so we 
have that $k_c^{out}\sim (N-n_c)\langle k\rangle$ and $k_c^{in}\sim n_c\langle k\rangle$. By definition of the linking probabilities $p_{in}$ and $p_{out}$ we deduce that
\begin{equation}
\label{eq1}
p_{out}=\frac{k_{i}^{out}}{k_c^{out}}\sim \frac{k_{i}^{out}}{(N-n_c)\langle k\rangle},
\end{equation}
and
\begin{equation}
\label{eq2}
p_{in}=\frac{k_{i}^{in}}{k_c^{in}}\sim \frac{k_{i}^{in}}{n_c\langle k\rangle}.
\end{equation}
In this way, the condition for the existence of communities $p_{in}>p_{out}$ becomes
\begin{equation}
\label{eq3}
\frac{k_{i}^{in}}{n_c\langle k\rangle}>\frac{k_{i}^{out}}{(N-n_c)\langle k\rangle},
\end{equation}
from which we get 
\begin{equation}
\label{eq4}
k_{i}^{in}>\frac{n_ck_{i}^{out}}{N-n_c}.
\end{equation}
On the other hand, by definition we have that
\begin{equation}
\label{eq5}
\mu=\frac{k_{i}^{out}}{k_{i}^{in}+k_{i}^{out}}.
\end{equation}
By comparing Eq.~\ref{eq5} with Eq.~\ref{eq4} we obtain the desired condition on $\mu$
\begin{equation}
\label{eq6}
\mu<\frac{N-n_c}{N}.
\end{equation}

The condition expressed in Eq.~\ref{eq6} is general, and applies to any version of the planted $\ell$-partition model. 
When communities are different in size, the upper bound on $\mu$ depends on the specific community at hand. However, if $n_c^{max}$ is the size of the largest
community, we can safely assume that, whenever $\mu<(N-n_c^{max})/N$, all communities are well defined.
In the GN benchmark, where $n_c=32$ and $128$, the condition becomes $\mu<3/4$. This is interesting,
as in most works using the GN benchmark, one usually assumes that communities are there as long as $\mu<1/2$, whereas 
they are not well defined for $\mu>1/2$. Instead, we see that communities are there, at least in principle, up until $\mu=3/4$. 
However, we stress that, even if communities are there, methods may be unable to detect them. The reason is that, due to fluctuations
in the distribution of links in the graphs, already before the limit imposed by the planted partition model it may be impossible
to detect the communities and the model graphs may look similar to random graphs. This issue of the actual significance of communities
and their detectability {\it a priori} is very important and has been recently discussed in the literature~\cite{reichardt08,bianconi08b,lancichinetti09c}.  
We notice that, on large networks, when $n_c\ll N$, the limit value of $\mu$ below which communities are defined approaches $1$. In our tests with the 
LFR benchmark, we will often be in this regime.

\section{Comparing partitions}
\label{sec10}

Testing an algorithm on any graph with built-in community structure also implies defining a quantitative criterion to 
estimate the goodness of the answer given by the algorithm as compared to the real answer that is expected. This can be done by 
using suitable similarity measures. For reviews of similarity measures see Refs.~\cite{fortunato09,meila07,traud08}. In the first tests
of community detection algorithms, one used a measure called {\it fraction of correctly identified nodes}, introduced by Girvan and Newman~\cite{girvan02}.
However, it is not well defined in some cases (e. g. when a detected community is a merger of two or more ``real'' communities), so in the last years
other measures have been used. In particular, measures borrowed from information theory have proved to be reliable.

To evaluate the Shannon information content~\cite{mackay03} of a partition,
one starts by considering the community assignments $\{x_i\}$ and $\{y_i\}$, where $x_i$ and $y_i$ indicate the cluster labels of 
vertex $i$ in partition ${\cal X}$ and ${\cal Y}$, respectively. One assumes that the labels $x$ and $y$ are values of two random
variables $X$ and $Y$, with joint distribution $P(x,y)=P(X=x,Y=y)=n_{xy}/n$, which implies that $P(x)=P(X=x)=n_x^X/n$ 
and $P(y)=P(Y=y)=n_y^Y/n$, where $n_x^X$, $n_y^Y$ and $n_{xy}$ are the sizes of the clusters labeled by $x$, $y$ and of their overlap, respectively. 
The {\it mutual information} $I(X,Y)$ of two random variables is defined as
\begin{equation}
I(X,Y)=\sum_{x}\sum_{y}P(x,y)\log\frac{P(x,y)}{P(x)P(y)}.
\label{eqstinf8}
\end{equation}
The measure $I(X,Y)$ tells how much we learn about $X$ if we know $Y$, and viceversa.
Actually $I(X,Y)=H(X)-H(X|Y)$, where $H(X)=-\sum_xP(x)\log P(x)$ is the Shannon entropy of $X$ and $H(X|Y)=-\sum_{x,y}P(x,y)\log P(x|y)$
is the conditional entropy of $X$ given $Y$. The mutual information is not ideal as a similarity measure: in fact, given a partition
${\cal X}$, all partitions derived from ${\cal X}$ by further partitioning (some of) its clusters would all have 
the same mutual information with ${\cal X}$, even though they could be very different from each other. In this case the mutual information
would simply equal the entropy $H(X)$, because the conditional entropy would be systematically zero. To avoid that, Danon et al.
adopted the {\it normalized mutual information}~\cite{danon05}
\begin{equation}
I_{norm}({\cal X}, {\cal Y})=\frac{2I(X,Y)}{H(X)+H(Y)},
\label{eqt08}
\end{equation}
which equals $1$ if the partitions are identical, whereas it has an expected value of $0$ if the partitions are independent.
The normalized mutual information is currently very often used in tests of community detection algorithms.
We have recently proposed a definition of the measure to evaluate
the similarity of {\it covers}, i. e. of divisions of the network in overlapping communities, which one needs for the tests of Section~\ref{sec4.4}. 
The details can be found in the Appendix of Ref.~\cite{lancichinetti09}. We stress that our definition is not a proper extension
of the normalized mutual information, in the sense that it does not recover exactly the same value of the original measure for the comparison of proper partitions
without overlap, even though the values are close. For consistency we used our definition in all tests, although in the tests involving benchmarks without overlapping 
communities the classic expression of Eq.~\ref{eqt08} could be used. For this reason, we warn that in the plots showing the performance of the algorithms on the
GN benchmark, the curves are not identical to those already seen in previous papers (for, e. g., modularity-based methods), where Eq.~\ref{eqt08} was used, although they are rather close. 

\section{The algorithms}
\label{sec2}

We have tested a wide spectrum of community detection methods. In some cases the software to implement the algorithms was publicly available,
in other cases the original developers have let us use their own code, otherwise we have created the software on our own. 
We wanted to have a representative subset of algorithms, that exploit some of the most interesting 
ideas and techniques that have been developed over the last years. Obviously we could not by any means perform an analysis 
of all existing techniques, as their number is huge. Some of them were excluded a priori, if particularly slow, as our tests
involve graphs with a few thousand nodes, which old methods are unable
to handle. On the other hand, the code to create the LFR benchmark is freely available~\cite{note1} 
and scholars are welcome to test their algorithms on it and compare their performance with
that of the algorithms analyzed here. Here is the list of the algorithms we considered.
\begin{itemize}
\item{{\it Algorithm of Girvan and Newman}~\cite{girvan02,newman04b}. It is the first algorithm of the modern age of community detection in graphs.
It is a hierarchical divisive algorithm, in which links are iteratively removed based on the value of their betweenness, which expresses
the number of shortest paths between pairs of nodes that pass through the link. In its most popular implementation, the procedure
of link removal ends when the modularity of the resulting partition reaches a maximum. The modularity of Newman and Girvan
is a well known quality function that estimates the goodness of a partition based on the comparison between the graph at hand and a null model,
which is a class of random graphs with the same expected degree sequence of the original graph. The algorithm has a complexity $O(N^3)$ on a sparse graph.
In the following we will refer to it as GN.}
\item{{\it Fast greedy modularity optimization by Clauset, Newman and Moore}~\cite{clauset04}. This method is essentially a fast implementation of 
a previous technique proposed by Newman~\cite{newman04c}. Starting from a set of isolated nodes, the links of the original graph are iteratively 
added such to produce the largest possible increase of the modularity of Newman and Girvan at each step. The fast version of Clauset, Newman and Moore, which
uses more efficient data structures, has a complexity of $O(N\log^2N)$ on sparse graphs.}
\item{{\it Exhaustive modularity optimization via simulated annealing}~\cite{guimera04,massen05,medus05,guimera05}. The goal is the same as in the previous
algorithm, but the precision of the final estimate of the maximum is far higher, due to the exhaustive optimization, at the expense of the 
computational speed. The latter cannot be expressed in closed form, as in the cases above, as it depends on the parameters used for the optimization. We will
stick to the procedure used by Guimer\'a and Amaral~\cite{guimera05}.}
\item{{\it Fast modularity optimization by Blondel et al.}~\cite{blondel08}. This is a multistep technique based on a local optimization of Newman-Girvan
modularity in the neighborhood of each node. After a partition is identified in this way, communities are replaced by supernodes, yielding a smaller weighted 
network. The procedure is then iterated, until modularity (which is always computed with respect to the original graph) does not increase any further. This
method offers a fair compromise between the accuracy of the estimate of the modularity maximum, which is better than that delivered by greedy techniques
like the one by Clauset et al. above, and computational complexity, which is essentially linear in the number of links of the graph.}
\item{{\it Algorithm by Radicchi et al.}~\cite{radicchi04}. This algorithm is in the spirit of that by Girvan and Newman above. In fact, it is 
a divisive hierarchical method, where links are iteratively removed based on the value of their edge clustering coefficient, which is defined as the ratio between 
the number of loops based on the link and the largest possible number of loops that can be based on the link. The edge clustering coefficient is 
a local measure, so its computation is not so heavy as that of edge betweenness, which yields a significant improvement in the complexity of the algorithm,
which is $O(N^2)$ on a sparse graph. Another major difference from the GN algorithm is the stopping criterion of the procedure, which depends on the properties of the
communities themselves and not on the values of a quality function like modularity. Radicchi et al. considered two types of communities: {\it strong} communities
are groups of nodes such that the internal degree of each node exceeds its external degree; {\it weak} communities are groups of nodes such that
the total internal degree of the nodes of the group exceeds their total external degree.}
\item{{\it Cfinder}~\cite{palla05}. This is a local algorithm proposed by Palla et al.
that looks for communities that may overlap, i.e. share nodes. It was the first paper
in the physics literature on community detection to address this problem, which is important in many systems like, e. g., social networks. Communities are
defined as the largest possible subgraphs that can be explored by rolling $k$-cliques across the network, where a $k$-clique rolls by rotating about 
any of its component $(k-1)$-cliques (which are links when $k=3$). The complexity of this procedure can be high, 
as the computational time needed to find all $k$-cliques of a graph is an exponentially growing function of the graph size~\cite{bron73}, but in practical applications
the method is rather fast, enabling one to analyze systems with up to $10^5$ nodes.}
\item{{\it Markov Cluster Algorithm}~\cite{vandongen00}. This is an algorithm developed by S. Van Dongen, which simulates a peculiar diffusion process on the graph.
One starts from the right stochastic matrix (or diffusion matrix) of the graph, which is obtained from the adjacency matrix of the original graph by 
dividing the elements of each row by their sum. Then one computes an integer power of this matrix (usually the square), which yields the probability matrix of 
a random walk after a number of steps equal to the number of powers of the right stochastic matrix considered. This step is called expansion.
Next, each element of the matrix is 
raised to some power $\alpha$, in order to enhance (artificially) the probability of the walker to be trapped within a community. This step is called inflation.
The expansion and inflation steps are iterated until one obtains the adjacency matrix of a forest (i. e. a disconnected tree), whose components are the communities.
This method, widely used in bioinformatics, is strongly dependent on the choice of the parameter $\alpha$. Its complexity can be lowered to $O(Nk^2)$ if, after each 
inflation steps, only the $k$ largest elements of the resulting matrix are kept, whereas the others are set to zero. In the following we will refer to the method as MCL.}
\item{{\it Structural algorithm by Rosvall and Bergstrom}~\cite{rosvall07}. Here the problem of finding the best cluster structure of a graph is turned into
the problem of optimally compressing the information on the structure of the graph, so that one can recover as closely as possible the original structure when the compressed
information is decoded. This is achieved by computing the minimum of a function which expresses 
the best tradeoff between the minimal conditional information between the original and the compressed information (maximal faithfulness to the original information) 
and the maximal compression (least possible information to transmit). The optimization of the function is carried out via simulated annealing, which makes the algorithm
quite slow, although one could always go for a faster and less accurate optimization.
In the following we will refer to the method as Infomod.}
\item{{\it Dynamic algorithm by Rosvall and Bergstrom}~\cite{rosvall08}. This technique is based on the same principle as the previous one. The difference is that
before one was compressing the information on the structure of the graph, here one wishes to compress the information of a dynamic process
taking place on the graph, namely a random walk. The optimal compression is achieved again by optimizing a quality function, which is the 
Minimum Description Length~\cite{rissanen78,grunwald05} of the random walk. Such optimization can be carried out rather quickly with a combination of 
greedy search and simulated annealing. In the following we will refer to the method as Infomap.}
\item{{\it Spectral algorithm by Donetti and Mu{\~n}oz}~\cite{donetti04}. This is a method based on spectral properties of the graph. The idea is that
eigenvector components corresponding to nodes in the same community should have similar values, if communities are well identified. Donetti and Mu\~noz
focused on the eigenvectors of the Laplacian matrix. They considered a limited number of eigenvectors, say $g$, and represented each node of the graph
as a geometric point in an Euclidean $g$-dimensional space, whose coordinates are the eigenvector components corresponding to the node. The points are then
grouped with traditional hierarchical clustering techniques. Of the resulting partitions, one picks the one that maximizes the modularity
by Newman and Girvan. The method is rather quick when only a few eigenvectors are computed, which is usually the case, as this can be done via the
Lanczos method~\cite{lanczos50}. In the following we will refer to the method as DM.}
\item{{\it Expectation-maximization algorithm by Newman and Leicht}~\cite{newman07}. Here Bayesian inference is used to deduce the best fit of 
a given model to the data represented by the actual graph structure. The goodness of the fit is expressed by a likelihood
that is maximized by means of the expectation-maximization technique~\cite{dempster77}. This leads to a system of self-consistent equations, that can
be solved by iteration starting from suitable initial conditions. The equations can be solved rather quickly and fairly large systems can be analyzed in this way
(up until $10^6$ nodes).
A nice feature of the method is that it finds the most relevant group structure of 
the graph, whether the groups are communities or not (in graphs with multipartite structure the classes are rather anti-communities, as 
there are very few links inside the groups). A drawback of the method is the fact that one needs to feed the number of groups, which is usually not
known {\it a priori}. In the following we will refer to the method as EM.}
\item{{\it Potts model approach by Ronhovde and Nussinov}~\cite{ronhovde08b}. This method is based on the minimization of the Hamiltonian of a Potts-like spin model,
where the spin state represents the membership of the node in a given community. A resolution parameter enables one to span several community scales, from
very small to very large communities. The relevant scales are identified by checking for the stability of the partitions obtained for given values
of the resolution parameter. This is done by computing the similarity of partitions obtained for the same resolution parameter but starting from different
initial conditions. Peaks in the similarity spectrum correspond to stable/relevant partitions. The method is rather fast, its complexity is slightly superlinear in the
number of links of the graph. In the following we will refer to the method as RN.}
\end{itemize}

\section{Tests on the GN benchmark}
\label{sec3}

\begin{figure} [ht]
\centering
\includegraphics[width=\columnwidth]{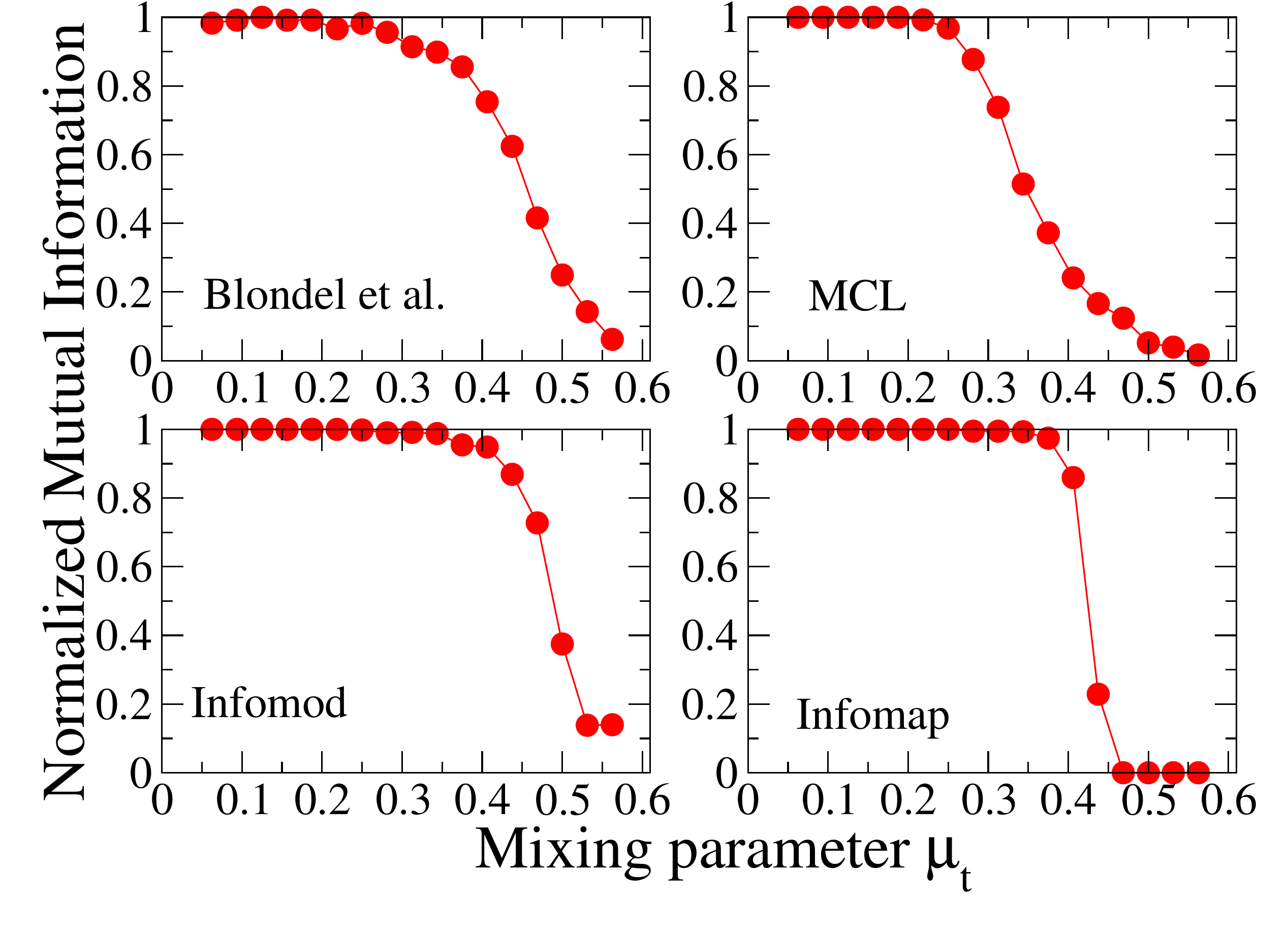} 
\includegraphics[width=\columnwidth]{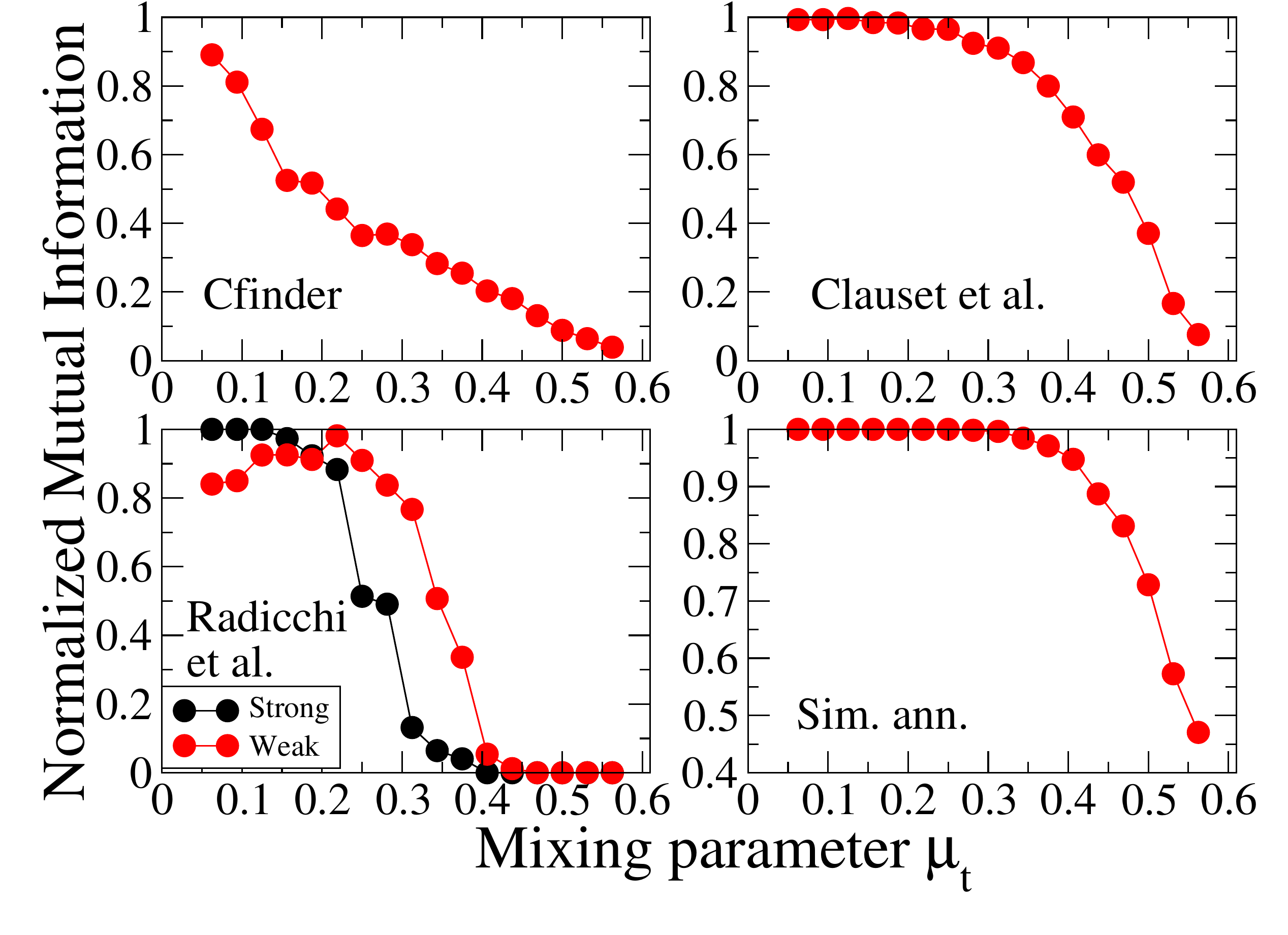} 
\includegraphics[width=\columnwidth]{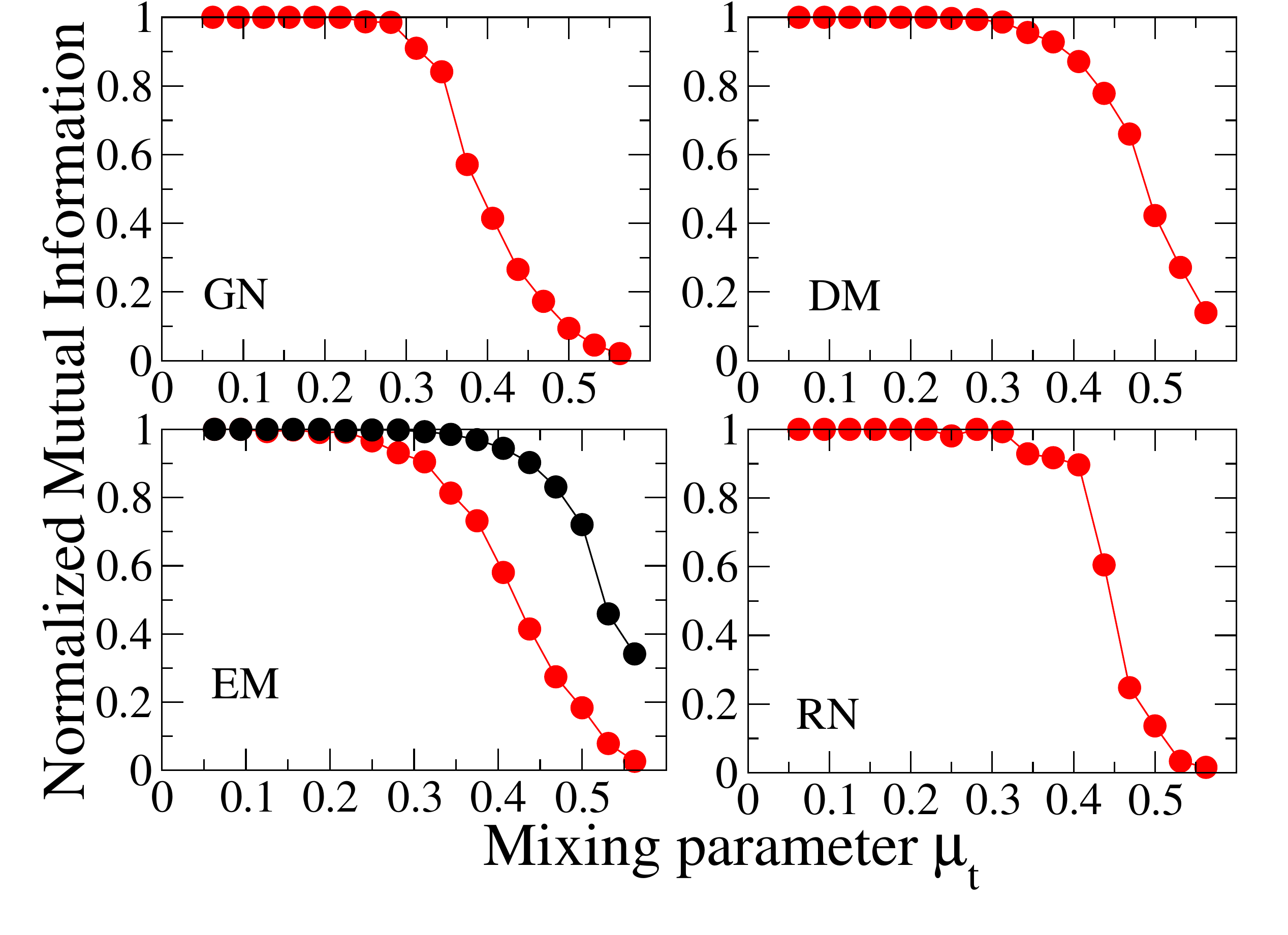} 
\caption {\label{GN} Tests of the algorithms on the GN benchmark.} 
\end{figure}
We begin by showing the performance of the algorithms on the GN benchmark. As we have explained in 
Section~\ref{sec1}, for the GN benchmark communities are well defined (in principle) up until a value $3/4=0.75$ for the mixing parameter.
We will indicate the mixing parameter with the symbol $\mu_t$ to mean that we refer to topology. In Section~\ref{sec4.2}
we will focus instead on the mixing parameter $\mu_w$, which considers the weights of the links. 
In Fig.~\ref{GN} we show the results of our analysis.
Each point of every curve corresponds to an average over $100$ realizations of the benchmark.  
For the algorithms by Radicchi et al. and by Newman and Leicht (EM),
we have put two curves instead of one (likewise in Section~\ref{sec4.1}). 
In the first case, we showed the outcome of the method when one uses both possible stopping
criteria, corresponding to a partition consisting of strong (black curve) and weak (red curve) communities, respectively. In the 
case of the EM method, we show the curves delivered by the iterative solution of the EM equations when one starts from a random partition 
(red), and from the planted partition of the benchmark (black curve). As one can see, results are different in these cases, even if they are
solutions of the same equation. This shows how sensitive the solution is to the choice of the initial condition. Moreover,
the maximum likelihood achieved when one makes the ``intelligent guess'' of the real partition is higher compared to the 
maximum likelihood obtained starting from a random partition. This indicates that the greedy approach to the solution of the 
EM equations suggested by Newman and Leicht is not an efficient way to maximize the likelihood, as one may expect.

Most methods perform rather well, although all of them start to fail 
much earlier than the expected threshold of $3/4$. The Cfinder fails to detect the communities even
when $\mu_t\sim 0$, when they are very well identified. This is due to the fact that, even when $\mu_t$ is small, the probe clique
that explores the system manages to pass from one group to the other and yields much larger groups, often spanning the whole graph.
The method by Radicchi et al. does not have a remarkable performance either, as it
also starts to fail for low values of $\mu_t$, although it does better than the Cfinder. The MCL 
is better than the method by Radicchi et al., but is outperformed by modularity-based methods 
(simulated annealing, Clauset et al., Blondel et al.), which generally 
do quite well on the GN benchmark, something that was already known from the literature. 
The DM and RN methods have a comparable performance as the exhaustive optimization of modularity via simulated annealing.
The GN algorithm performs about as well as the MCL.
Both methods by Rosvall and Bergstrom have a good performance. In fact, up until $\mu_t\sim 0.4$, they always guess the planted
partition in four clusters. 

\section{Tests on the LFR benchmark}
\label{sec4}

In this section we will present the tests on the LFR benchmark. For a thorough analysis, we have considered
various versions of the benchmark, in which links can have or not weights and/or direction. We have also examined the 
version which allows for community overlaps. In each test, we have averaged the value of the normalized mutual information
over $100$ realizations for each value of the mixing parameter.

\subsection{Undirected and unweighted graphs}
\label{sec4.1}

\begin{figure} [ht]
\centering
\includegraphics[width=\columnwidth]{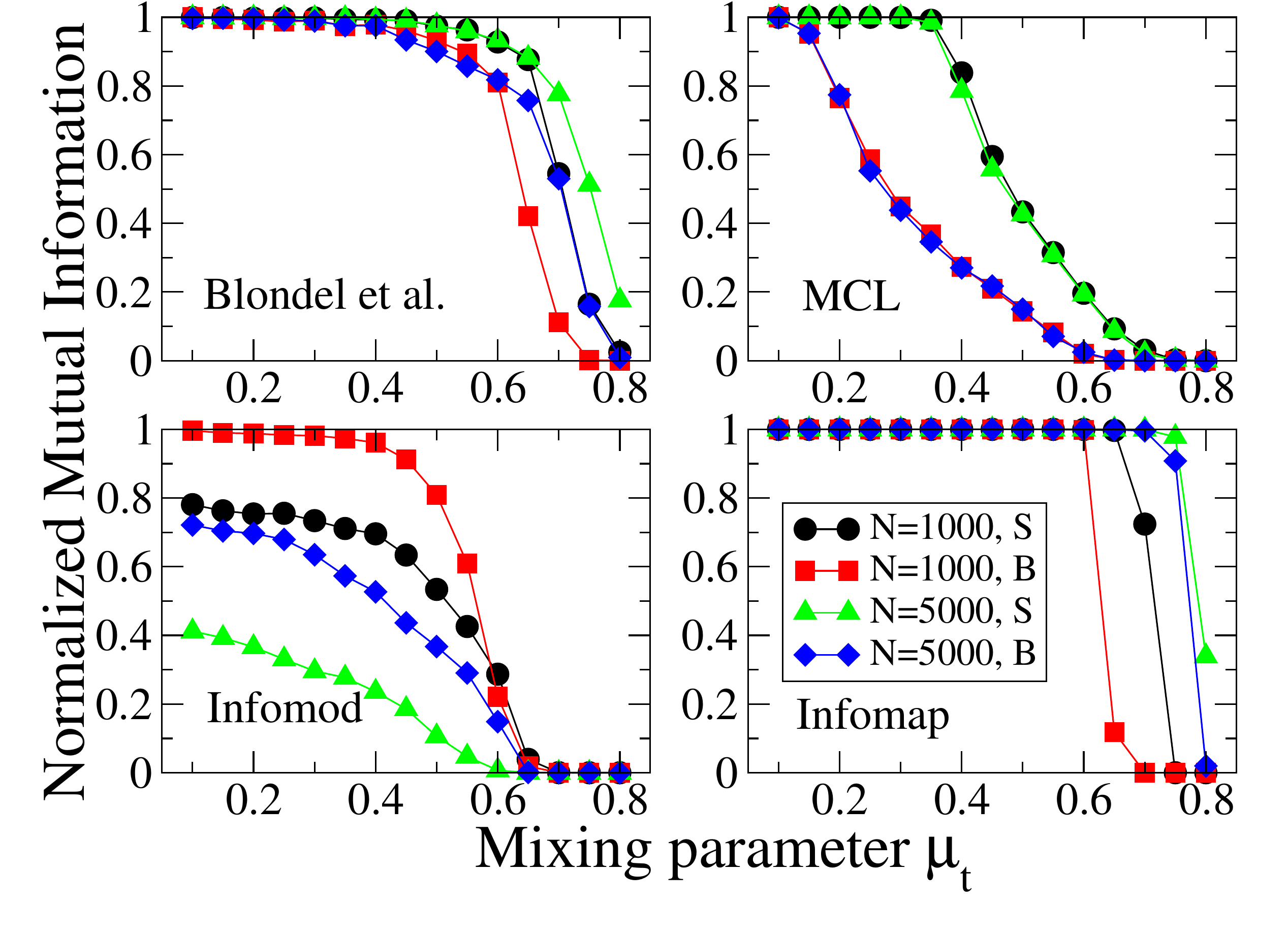} 
\includegraphics[width=\columnwidth]{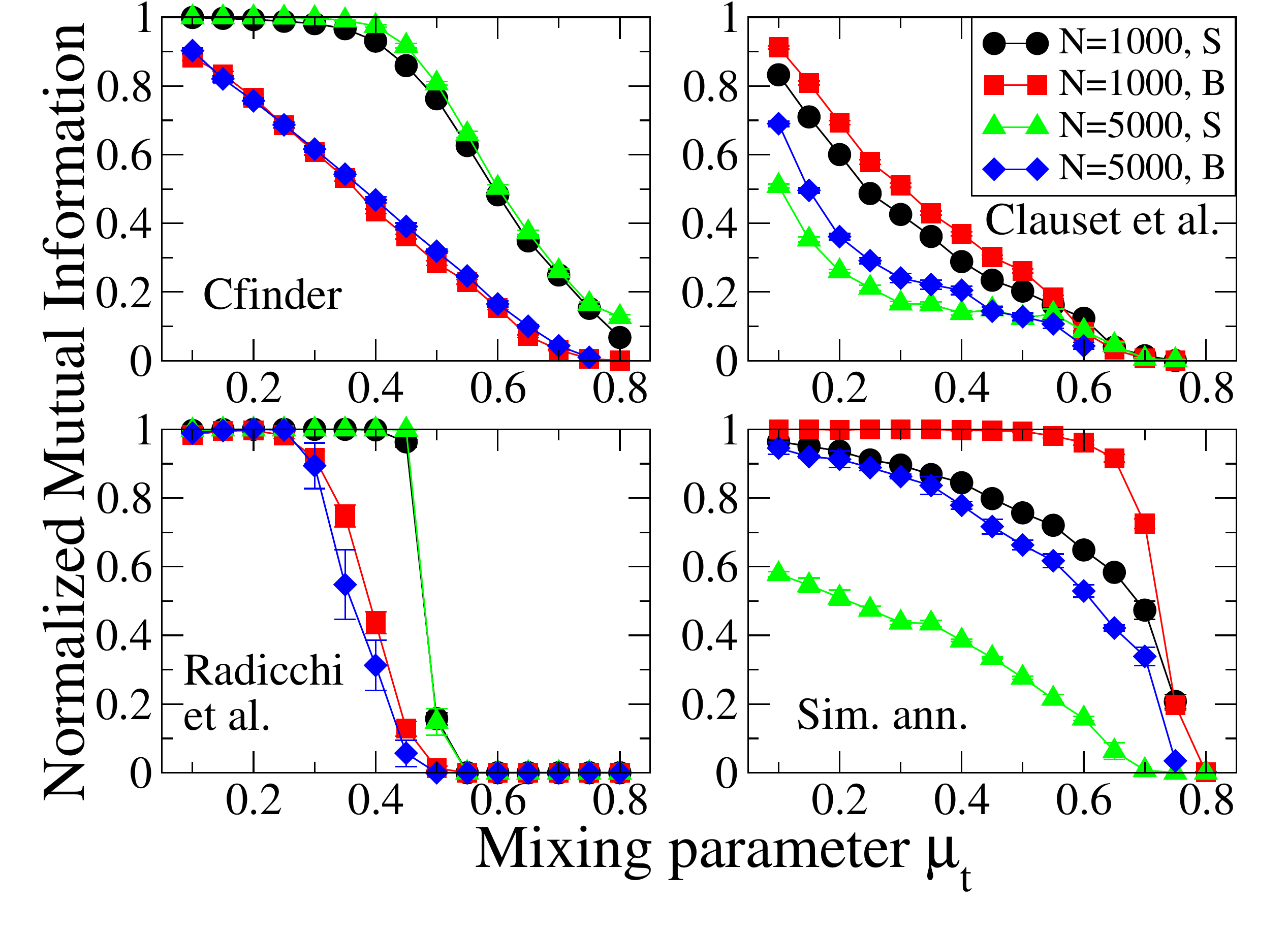} 
\includegraphics[width=\columnwidth]{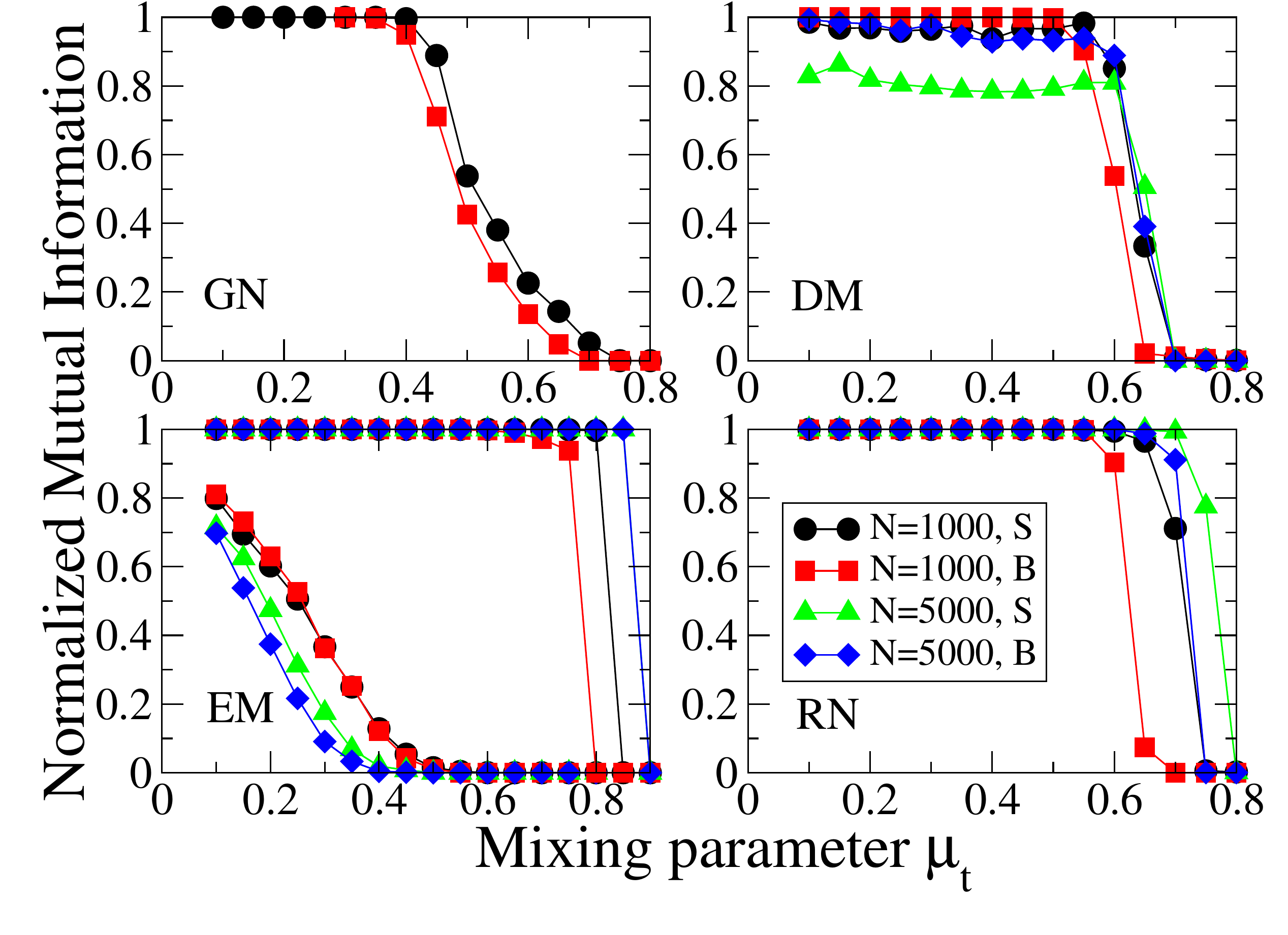} 
\caption {\label{LFR} Tests of the algorithms on the LFR benchmark with undirected and unweighted links.} 
\end{figure}

The plots of Fig.~\ref{LFR} illustrate the results of the analysis. The following input parameters
are the same for all benchmark graphs used here, as well as in Sections~\ref{sec4.3},
\ref{sec4.2} and \ref{sec4.4}: the average degree is $20$, the maximum degree $50$, the exponent of the degree
distribution is $-2$ and that of the community size distribution is $-1$.
In each plot, except for the GN and the EM algorithms, 
we show four curves, corresponding to two 
different network sizes ($1000$ and $5000$ nodes) and, for a given size, to two different ranges for the community sizes, indicated
by the letters $S$ and $B$: $S$ (stays for ``small'') means that communities have between $10$ and $50$ nodes, $B$ (stays for ``big'')
means that communities have between $20$ and $100$ nodes. For the GN algorithm we show only
the curves corresponding to the smaller network size, as it would have taken too long to accumulate enough statistics to 
present clean plots for networks of $5000$ nodes, due to the high computational complexity of the method. 
For the EM method we have plotted eight curves
as for each set of benchmark graphs we have considered the two outcomes of the algorithm corrsponding to the different choices of initial conditions
we have mentioned in the previous section, namely random (bottom curves) and planted partition (top curves). 
In this case, the difference in the performance of the algorithm
in the two cases is remarkable. The fact that, by starting from the planted partition, the final likelihood is 
actually higher as compared with a random start, as we have seen in the previous section, confirms
that the method has a great potential, if only one could find a better way to estimate the maximum likelihood than the greedy approach 
currently adopted.
Nevertheless we remind that the EM also has the big drawback to require as input the number of groups to be found, which is usually unknown in applications.

As a general remark, we see that the LFR benchmark enables one to discriminate the performances of the algorithms much better than the GN benchmark, as expected.
Modularity-based methods have a rather poor performance, which worsens for larger systems and smaller communities, due to the well known resolution limit
of the measure~\cite{fortunato07}.
The only exception is represented by the 
algorithm by Blondel et al., whose performance is very good, probably because the estimated modularity maximum is not a 
very good approximation of the real one, which is more likely found by simulated annealing. The Cfinder, the MCL and the method by
Radicchi et al. do not have impressive
performances either, and display a similar pattern, i.e. the performance is severely affected by the size of the communities (for larger communities
it gets worse, whereas for small communities it is decent), 
whereas it looks rather insensitive to the size of the network. The DM has a fair performance, but it gets worse 
if the network size increases. The same trend is shown by Infomod, where the performance worsens considerably with the increase of the network size.  
Infomap and RN have the best performances, with the same pattern with respect to the size of the network and of the communities: up to
values of $\mu_t\sim 1/2$ both methods are capable to derive the planted partition in the $100\%$ of cases. 

\begin{figure} [ht]
\centering
\includegraphics[width=\columnwidth]{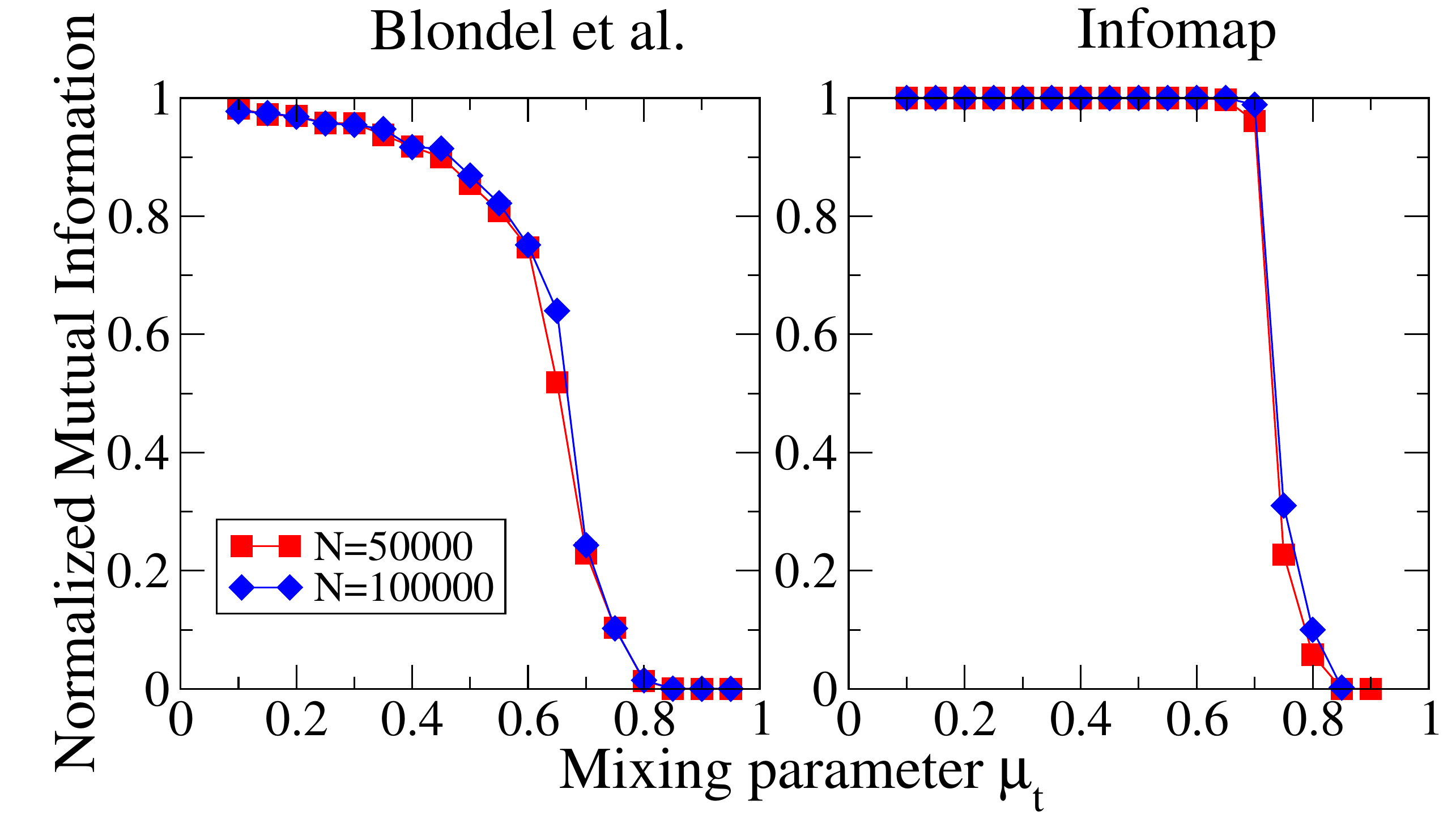} 
\caption {\label{LFRl} Tests of the algorithm by Blondel et al. and Infomap on large LFR benchmark graphs with undirected and unweighted links.} 
\end{figure}

We conclude that Infomap, the RN method and the method by Blondel et al. are the best performing algorithms on the LFR undirected and unweighted
benchmark. Since Infomap and the method by Blondel et al. are also very fast, essentially linear in the network size, we wonder how good their 
performance is on much larger graphs than those considered in Fig.~\ref{LFR}. For this reason we carried out another set of tests of these two algorithms 
on the LFR benchmark, by considering graphs with $50000$ and $100000$ nodes. We have done so also because in the tests that can be found in the 
literature on community detection one typically uses very small graphs, and the performance can change considerably on large graphs.
In Fig.~\ref{LFRl} we show the performance of the two methods. Due to the large network size, we decided to pick a broad range
of community sizes, from $20$ to $1000$ nodes. In this way, the heterogeneity of the community sizes is manifest. The maximum degree here was fixed to $200$.
Remarkably, the 
performance of the method by Blondel et al. is worse than on the smaller graphs of Fig.~\ref{LFR}, whereas that of Infomap is stable
and does not seem to be affected.

\subsection{Directed and unweighted graphs}
\label{sec4.3}

\begin{figure} [ht]
\centering
\includegraphics[width=\columnwidth]{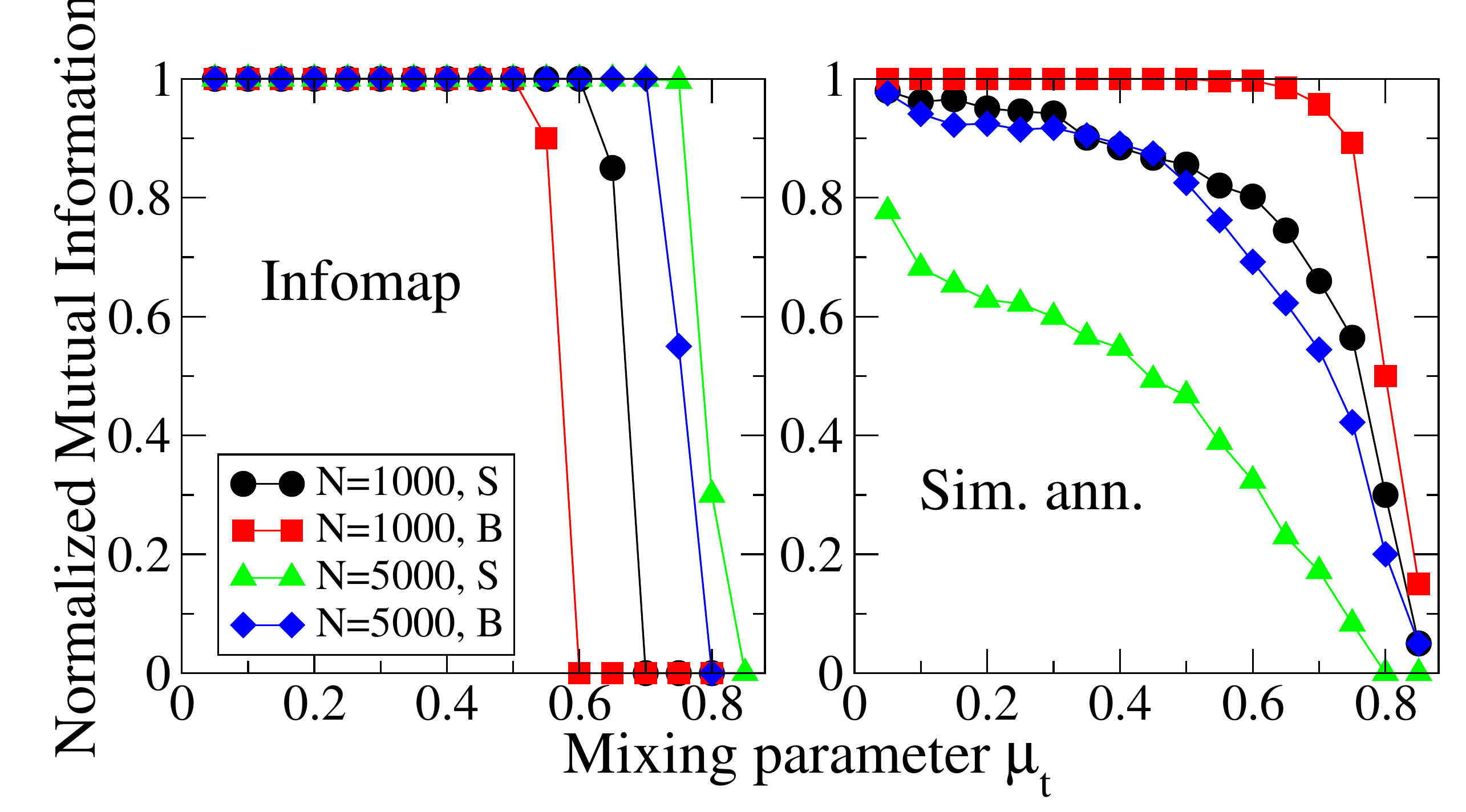} 
\caption {\label{LFRd} Tests of Infomap and of the exhaustive modularity optimization via simulated annealing 
on the LFR benchmark with directed and unweighted links.} 
\end{figure}

Directedness is an essential features of many real networks. Ignoring direction, as one often does or is forced to do, 
may reduce considerably the information that one can extract from the network structure. 
In particular, neglecting link directedness when looking for communities may lead to partial, or even misleading, results.
In the literature there has been no benchmark for directed graphs with communities for a long time. However, we have recently extended the  
LFR benchmark to directed networks~\cite{lancichinetti09b}, so we are in the position to 
evaluate the performance of community detection algorithms in this case. The presence of directed links is a serious obstacle
towards a generalization of an algorithm for community detection. Therefore, very few algorithms currently available are able to 
handle directed graphs. In the set of methods we consider here, only five can be used as well for directed networks: Clauset et al., simulated annealing for modularity, 
Cfinder, Infomap, EM. For some of the other algorithms one may think of possible extensions which are, at present, still missing.
The EM method, in its original definition of Ref.~\cite{newman07}, has actually problems to deal with directed graphs~\cite{ramasco08}.
We present here a comparison of the performances of two methods, exhaustive modularity optimization via simulated annealing and Infomap.
The results are in Fig.~\ref{LFRd}. Here the topological mixing parameter $\mu_t$ refers to the indegree of the nodes, which are distributed 
according to a power law as in the original undirected benchmark, while the outdegree is kept constant for all nodes, a choice made to avoid an unnecessary
proliferation of input parameters. Again, we considered two different network sizes and ranges for the community size, which are the same as those in Fig.~\ref{LFR}.
The other input parameters for the benchmark are the same that we have given in Section~\ref{sec4.1}.
As expected, modularity optimization shows the same limits that emerged in Fig.~\ref{LFR}. On the other hand, the 
performance of Infomap is still very good. 

\subsection{Undirected and weighted graphs}
\label{sec4.2}

In this section we focus on undirected graphs with weighted links. 
\begin{figure} [ht]
\centering
\includegraphics[width=\columnwidth]{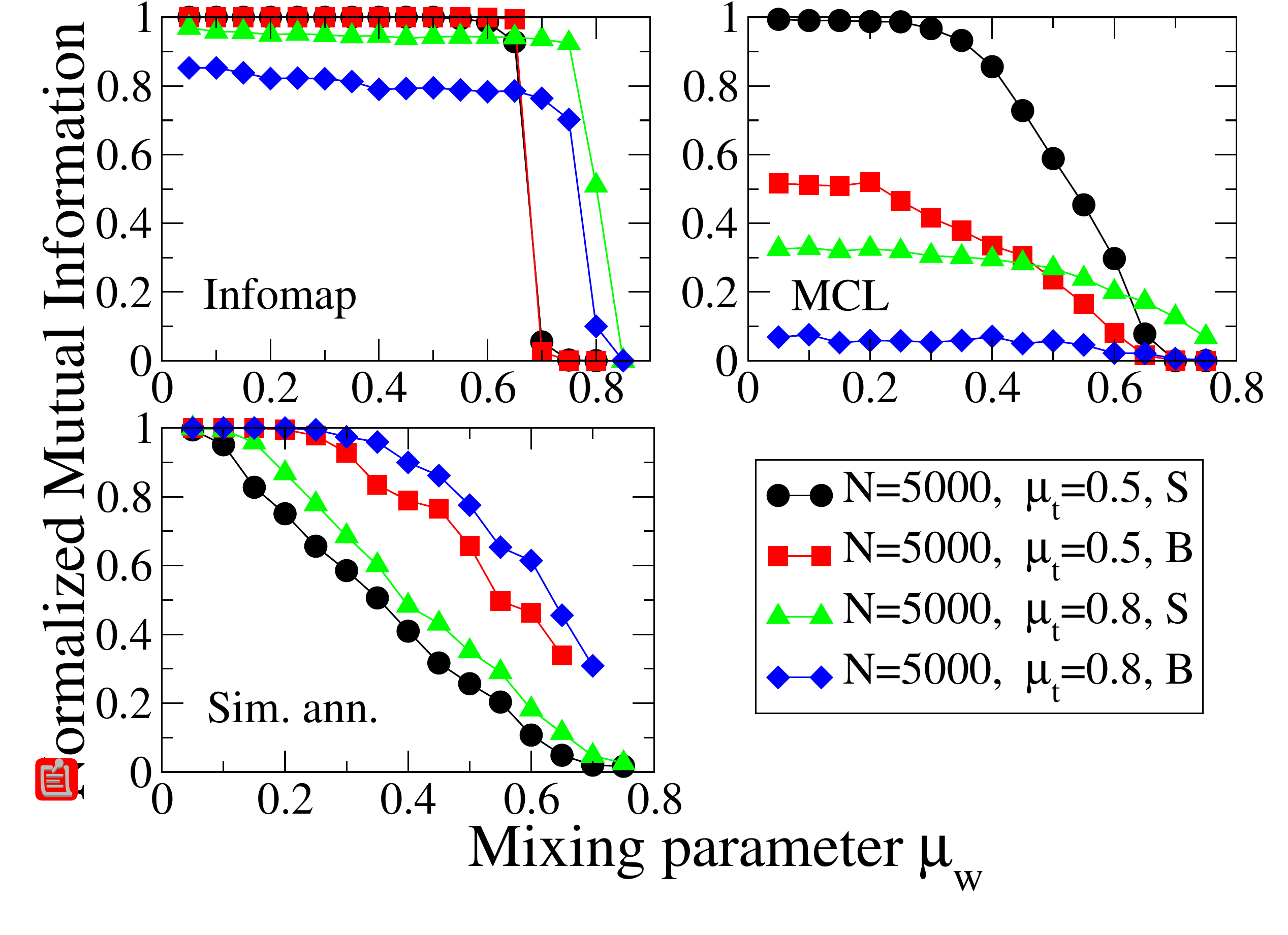} 
\caption {\label{LFRw} Tests of Infomap, MCL and of the exhaustive modularity optimization via simulated annealing 
on the LFR benchmark with undirected and weighted links.} 
\end{figure}
Weights are also precious sources of information~\cite{barrat04}.
Just as in the case of link directedness above, neglecting weights may imply a significant limitation of the information on a graph's properties,
concealing features of real systems which may be very important and not deducible from the mere topology. Ideally, one should 
exploit the information from both topology and weights for a reliable analysis of a network. The LFR benchmark has been extended to 
weighted graphs as well~\cite{lancichinetti09b}. Now there are two mixing parameters, one for topology, which is the same $\mu_t$
we have defined and used so far, and the other for the weights, $\mu_w$, which is the weighted counterpart of $\mu_t$, i. e. it expresses
the fraction of the strength of the node that lies on links connecting the node to the nodes outside its community, with respect to the 
total strength of the node. We remind that the strength of the node is the sum of the weights of its links. Moreover, 
there is an additional parameter, i. e. the exponent of the distribution for the strength: we have set it to $1.5$ for all realizations. All other parameters are the same
specified in Section~\ref{sec4.1}.
Since we wish to show the results of the
test on 2-dimensional plots, as we have done so far, we need to keep fixed one of the two parameters and study the dependence on the other.
Here we freeze the topological mixing parameter $\mu_t$ and study the dependence of the results on $\mu_w$, so that we see how the performance
of an algorithm varies when only the weights are redistributed, but the topology is fixed. The results are in Fig.~\ref{LFRw}, where we consider only
three methods: Infomap, MCL and exhaustive modularity optimization via simulated annealing. The other methods have no weighted counterpart
or the code for the weighted version was not available. In each plot we show four curves, corresponding to two choices for the topological 
mixing parameter $\mu_t$ and the two usual ranges of small (S) and big (B) communities that we have used so far. The network size is $5000$ nodes in each case.
The Infomap by Rosvall and Bergstrom has, once more, a remarkable performance, although it worsens if communities are topologically more mixed
(higher $\mu_t$) and larger in size (B). The MCL has a fair performance only in one case, for $\mu_t=0.5$ and small communities, whereas in the
other extreme of big topological mixture and big communities it fails for any value of $\mu_w$. Modularity optimization seems to be more
sensitive to the community size than to the other parameters.

\subsection{Undirected and unweighted graphs with overlapping communities}
\label{sec4.4}

The fact that communities in real systems often overlap has attracted a lot of attention in the last years, leading to the 
creation of new algorithms able to deal with this special circumstance, starting from the first work by Palla et al.~\cite{palla05}.
Meanwhile, a few methods have been developed~\cite{baumes05,zhang07,newman07,lancichinetti09,nepusz08,evans09,ahn09,gregory09}, but none of them
has been thoroughly tested, except on a bunch of specific networks taken from the real world. Indeed, 
there have been no suitable benchmark graphs with overlapping 
community structure, until recently~\cite{sawardecker09,lancichinetti09b}. In particular, the LFR benchmark has been extended to
the case of overlapping communities~\cite{lancichinetti09b}, and we use it here. Of our set of algorithms, only the Cfinder is able to find 
overlapping communities. In principle also the EM method assigns to each node the probability that it belongs to any community, but then one would need
a criterion to define which, among such probability values, is significant and shall be taken or is not significant and shall be neglected. For this reason
we report the results of tests carried out with the Cfinder only.
\begin{figure} [ht]
\centering
\includegraphics[width=\columnwidth]{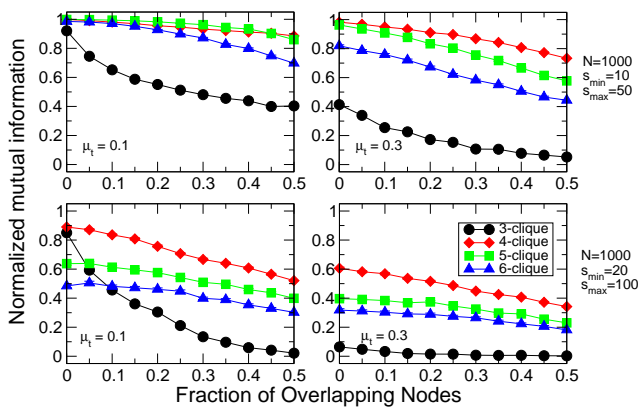} 
\caption {\label{LFRov1} Tests of the Cfinder on the LFR benchmark with undirected and unweighted links and overlapping communities.
The variable on the $x$-axis is the fraction of overlapping nodes. The networks have $1000$ nodes, the other parameters are $\tau_1=2$, $\tau_2=1$, 
$\langle k\rangle=20$ and $k_{max}=50$.} 
\end{figure}
\begin{figure} [ht]
\centering
\includegraphics[width=\columnwidth]{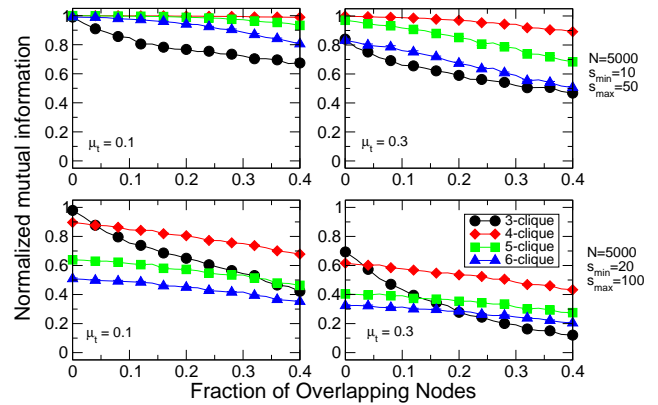} 
\caption {\label{LFRov2} Tests of the Cfinder on the LFR benchmark with undirected and unweighted links and overlapping communities.
The variable on the $x$-axis is the fraction of overlapping nodes. The networks have $5000$ nodes, the other parameters are the same 
used for the graphs of Fig.~\ref{LFRov1}.} 
\end{figure}

In Figs.~\ref{LFRov1} and \ref{LFRov2} we show the results.
The topological mixing parameter $\mu_t$ is fixed
and one varies the fraction of overlapping nodes between communities. We have run the Cfinder for different types of $k$-cliques ($k$ indicates
the number of nodes of the clique), with $k=3,4,5,6$. In general we notice that triangles ($k=3$) yield the worst performance,
whereas $4$- and $5$-cliques give better results. In the two top diagrams community sizes range between $10$ and $50$ nodes,
whereas in the bottom diagrams the range goes from $20$ to $100$ nodes. By comparing the diagrams in the
top with those in the bottom we see that the algorithm performs better
when communities are (on average) smaller. The networks
used to produce Fig.~\ref{LFRov1} consist of $1000$ nodes, whereas those of Fig.~\ref{LFRov2} consist of $5000$ nodes.
From the comparison of Fig.~\ref{LFRov1} with Fig.~\ref{LFRov2} we see that the algorithm performs better on networks of larger size.

\section{Tests on random graphs}
\label{sec5}

\begin{figure} [ht]
\centering
\includegraphics[width=\columnwidth]{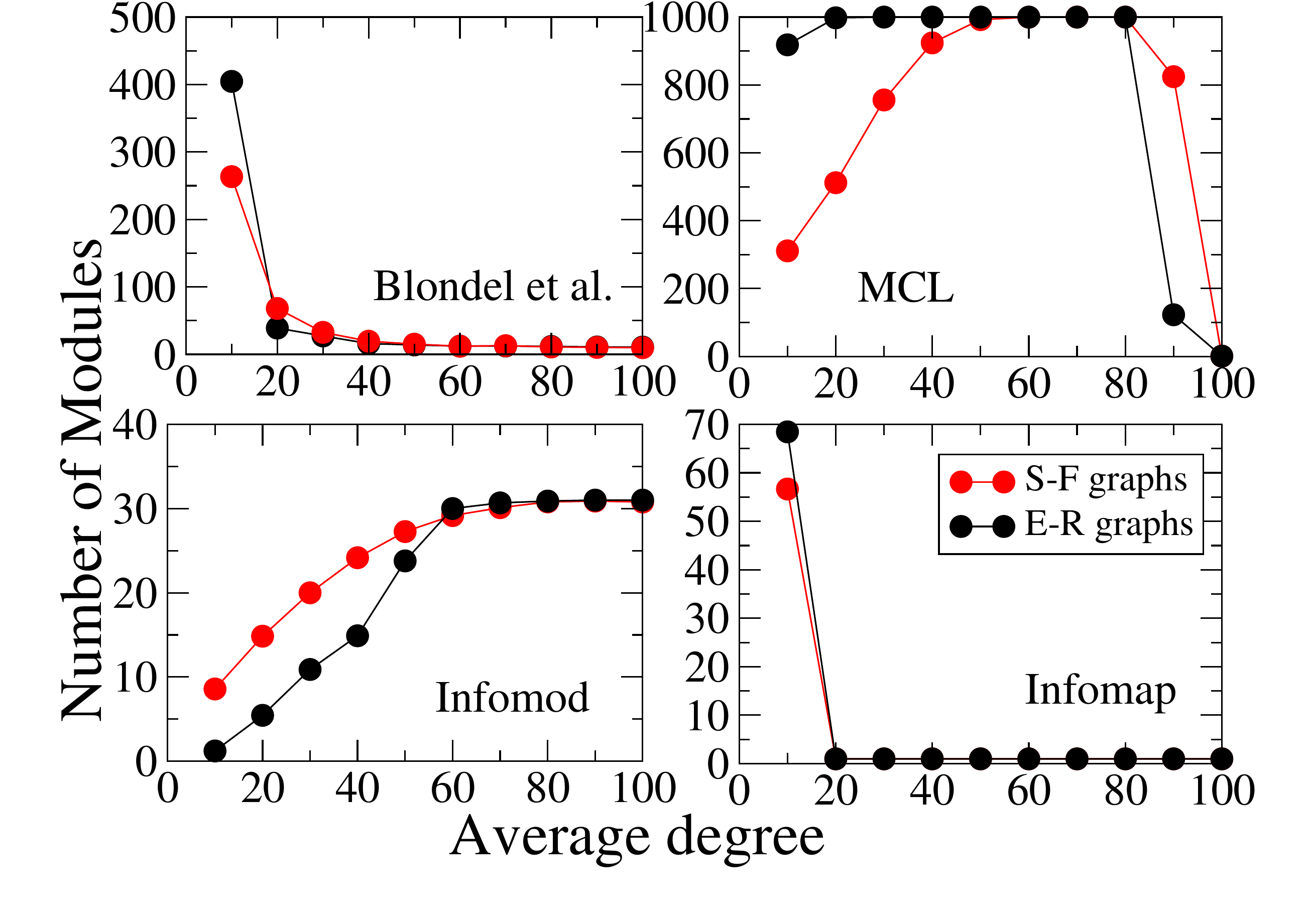} 
\includegraphics[width=\columnwidth]{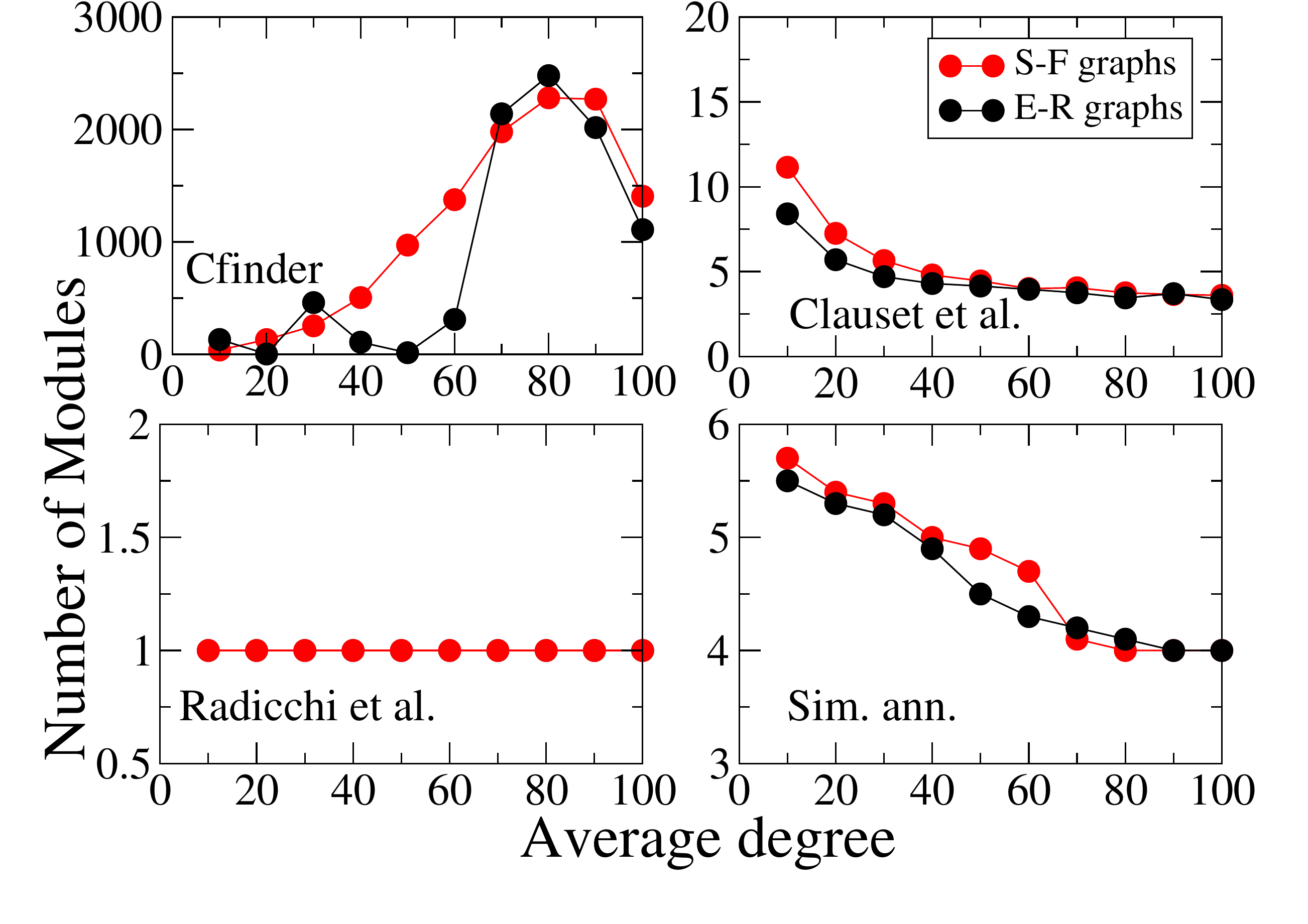} 
\includegraphics[width=\columnwidth]{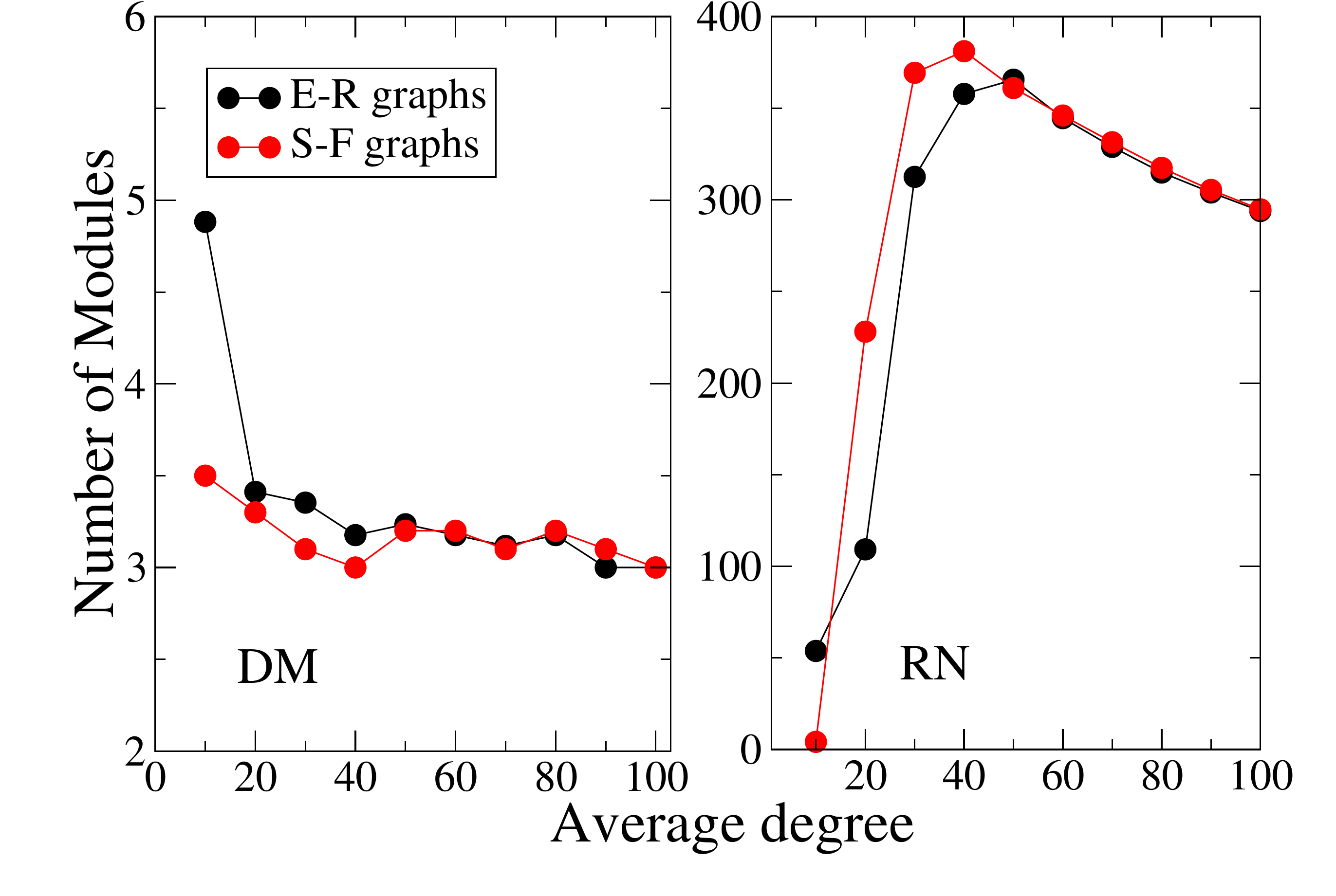} 
\caption {\label{rand} Tests of the algorithms on random graphs \'a la Erd\"os-R\'enyi (E-R) and scale free (S-F) random graphs.} 
\end{figure}

An important test of community detection algorithms, usually ignored in the literature, consists in applying them to random graphs.
In random graphs, by definition, the linking probabilities of the nodes are independent of each other. In this way one does not expect
that there will be inhomogeneity in the density of links on the graphs, i. e. there should be no communities. Things are not that simple, though. 
It is certainly true that {\it on average} this is what happens. On the other hand, specific realizations of random graphs may display
pseudocommunities, i. e., clusters produced by fluctuations in the link density. This is why, for instance, the maximum modularity of partitions
in random graphs is not small~\cite{guimera04, reichardt06,reichardt06b,reichardt07}. However, a good method should distinguish between such
pseudocommunities and meaningful modules. This is why we still expect to find no communities in random graphs.  We considered two types of graphs: random graphs
\'a la Erd\"os-R\'enyi~\cite{erdos59}, which have a binomial degree distribution, and random graphs with power law degree distributions (scale free). The latter have been
built via the configuration model~\cite{molloy95}, starting from a fixed degree sequence for the nodes obeying the predefinite power law distribution. The exponent of the
distribution is $-2$, the maximum degree was fixed to $200$. The size of all graphs, Erd\"os-R\'enyi and scale free, is fixed to $1000$ nodes.
In Fig.~\ref{rand} we show the number of modules found by various algorithms as a function of the average degree of the graph. Each point corresponds to an average over
$100$ graph realizations. We do not show the results of the EM method, because the number of modules must be given by input, and of the 
GN algorithm because it is too slow to be used for the analysis. 

The best performance is that of the method by Radicchi et al., which always finds a single cluster comprising all nodes. Another reasonable answer is
to find as many clusters as there are nodes, like the MCL. Here, however, the answer depends on the average degree $\langle k\rangle$ of the graph: if 
$\langle k\rangle$ is very low or very large the number of modules is smaller than $1000$, i.e. the method finds small groups of nodes.
This is particularly evident for scale free graphs. Modularity-based methods, like Clauset et al., the exhaustive optimization via simulated annealing, and
the algorithm by Blondel et al. are not so good, as they always find a few clusters, even in the limit of large $\langle k\rangle$: 
this is actually well known~\cite{reichardt06b}. 
This is also the case for the DM method, which performs a sort of modularity optimization, on the restricted set of partitions delivered 
by hierarchical clustering. Infomod and the RN method find non-trivial partitions for any value of $\langle k\rangle$. The Cfinder finds a single
module for very low values of $\langle k\rangle$ and then a rapidly rising number of modules as $\langle k\rangle$ increases. Since the modules
are strongly overlapping in this case, they may exceed the number of nodes, as we see from the plot. Instead, Infomap always finds a single 
module comprising all nodes, except when $\langle k\rangle$ is low.

\section{Summary}
\label{sec6}

We have carried out a comparative analysis of the performances of algorithms for community detection on various graphs: the GN and LFR benchmarks
and random graphs. Link direction, weights and the possibility for communities to overlap have been taken into account in dedicated tests.
We conclude that the Infomap method by Rosvall and Bergstrom~\cite{rosvall08} is the best performing on the set of benchmarks we have examined here. 
In particular, its results on the LFR benchmark graphs, which are much more 
difficult to examine than the GN benchmark graphs, 
as clearly shown by Figs.~\ref{GN} and \ref{LFR}, are encouraging about the reliability of the method in applications to real graphs.
Among the other things, the method can be applied to weighted and directed graphs as well, with excellent 
performances, so it has a large spectrum of potential applications.
The algorithms by Blondel et al.~\cite{blondel08} and by Ronhovde and Nussinov (RN)~\cite{ronhovde08b} 
also look very good from our analysis and could be used as well. In fact, 
for a study of the community structure in real graphs, one could think of using all three methods, to be able to extract some
algorithm-independent information. Furthermore, as we have seen in Section~\ref{sec2}, these methods have a low computational
complexity, so one could use them on graphs with millions of nodes and links. On the other hand, the algorithms are not able to 
account for overlapping communities, so they need to be properly refined to deal with this possibility, which is common in many real systems.
 
One may object that, despite the features planted in the LFR benchmark, i. e. the fat-tailed distributions of degree and 
community size, which are actually observed in real networks, our artificial graphs are still different from real systems. For instance,
the clustering coefficient~\cite{watts98} of the LFR benchmark is very low, due to the very small number of triangles, whereas real networks are characterized by 
many triangles and consequently a high clustering coefficient. On the one hand the GN benchmark also has very few triangles and low
clustering coefficient (the LFR benchmark is just a generalization of the GN benchmark), nevertheless people have used it 
extensively for testing algorithms. On the other hand, nothing forbids to modify the building mechanism of the LFR benchmark so that 
it does include triangles. This is actually a potentially interesting improvement of the benchmark, that deserves some attention in the future. 

Another important remark is in order. Our whole analysis has made use of graphs with a ``flat'' community structure, without hierarchy.
Many real networks instead have a hierarchical community structure, with communities inside other communities. Good methods must be able
to understand when a network has no communities, a flat or a hierarchical community structure. For an analysis of this kind we would need 
hierarchical benchmarks. There is actually a hierarchical version of the GN benchmark~\cite{arenas06}, not yet one of the LFR benchmark, which is 
sorely needed. Methods to find communities in multipartite graphs have yet to be tested as well.

From all of the above it is clear that this manuscript does not ``kill'' the issue of the actual
efficiency and reliability of community detection methods. Our analysis represents a first step, 
but it is clear that much more needs to be done along these lines.

\begin{acknowledgments}

We thank L. Donetti, R. Lambiotte, F. Radicchi, P. Ronhovde and M. Rosvall for kindly providing their code.
We are also grateful to V. Latora, T. Nepusz, A. Pluchino, A. Rapisarda, M. Sales-Pardo, C. Wiggins for useful suggestions.
We gratefully acknowledge ICTeCollective, grant number 238597 of the European Commission.

\end{acknowledgments}

\end{document}